\documentclass{JINST}

 \usepackage{subfigure}
 \usepackage{graphics}
\usepackage{lscape} 
\usepackage{rotating}
\usepackage{amssymb}
\usepackage{amsmath}
\usepackage{epsfig}

\title{Organic Liquid TPCs for Neutrino Physics}
\author{J. V. Dawson and D. Kryn\\
Laboratoire Astroparticule et Cosmologie, \\
10 rue Alice Domon et L\'eonie Duquet, 75205 Paris, France \\
E-mail: \email{jaime.dawson@gmail.com}}

\abstract{
We present a new concept for anti-neutrino detection, an organic liquid TPC with a volume of the order of m$^3$ and an energy resolution of the order of 1\% at 3 MeV and a sub-cm spatial resolution.
}

\begin{document}

\section{Introduction}
The current state of the art of anti-neutrino detectors (for example Double Chooz\cite{DoubleChooz}, Daya Bay\cite{DayaBay} and Reno \cite{Reno}) observe anti-neutrinos via inverse beta decay in liquid scintillator. The signal consists of two time-separated events. The first event is composed of the positron track and subsequent annihilation gamma rays which multiply Compton scatter over a wide volume. The second event is produced by neutron capture either on either hydrogen (2.2 MeV gamma ray) or
gadolinium which releases multiple gamma rays with a total energy of $\sim$8 MeV. The gamma rays deposit their energy via multiple Compton scatters in the low-Z scintillators. 

To improve on these detectors, obvious factors are 1) the energy resolution, 2) the background rejection and 3) the reconstruction of the direction of the incoming anti-neutrinos.
\begin{enumerate}

\item Energy resolution for these reactor $\theta_{13}$ experiments was not critical and so none of the experiments have maximal photo-coverage.  Proposals to use reactor neutrinos to resolve the mass hierarchy indicate the need for good energy resolution ($\sim$3\%) in kiloton scale detectors.  However scaling up and increasing the photo-coverage of the current reactor experiments may not suffice to achieve the necessary energy resolution. Further ideas are required.

\item The major correlated background to the anti-neutrino signal comes from cosmogenically produced isotopes, in particular $^{9}$Li and $^{8}$He. These isotopes decay with long half-lives, and so are not obviously correlated with their parent muon, emitting an electron and a neutron ($\beta$-neutron). For the current scintillator experiments, there is no means of distinguishing between an electron and a positron on an event-by-event basis.  

\item The Double Chooz detectors are capable of determining the average direction of the incoming anti-neutrinos \cite{neutrinodirectionality}. But their spatial resolution is not good enough to provide an indication of the direction event by event. %

\end{enumerate}

The detector technology proposed to meet these three challenges could serve many purposes. A large volume detector could be used to determine the neutrino mass hierarchy, and perform anti-neutrino cartography of the Earth. We find other interesting capabilities such as Particle Identification, discriminating between alpha, proton recoil, single Beta, double Beta and multiple Compton interactions. We also find a variety of target materials, some of which contain Double Beta isotopes.  Small detectors could be used for measuring neutron fluxes, studying decays from cosmogenic isotopes, and in searches for exotic decays such as Neutrinoless Double Beta Decay and rare beta decays.

\section{The Hydrogenous TPC}
\begin{figure}
\begin{center}
  \includegraphics[width=10cm]{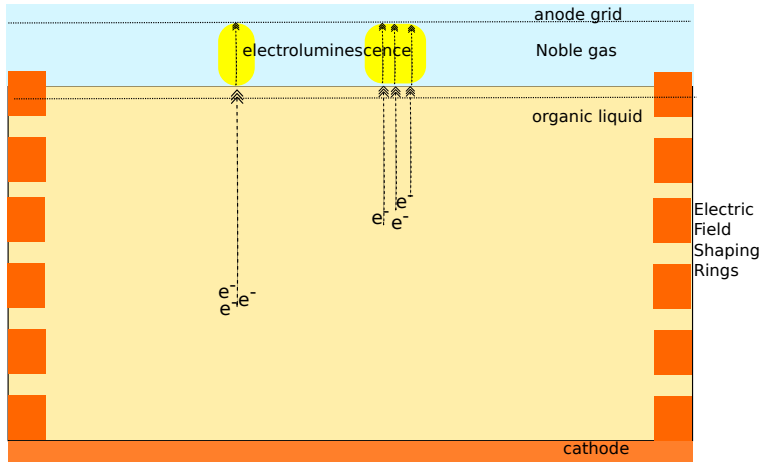} 
\caption{\small Simple cartoon of the proposed TPC. The electric field is controlled by anode grid, cathode, and electric field shaping rings. An additional grid under the liquid surface, used to control the electric field in the noble gas, is also shown.  Energy depositions in the organic liquid volume liberate electrons which drift under the applied electric field. Upon reaching the surface of the liquid, they are extracted in to the noble gas layer where electroluminescence occurs.}
\label{fig:TPC}
\end{center}
\end{figure}

We propose the novel concept of a hydrogenous TPC, a room temperature organic liquid covered with a layer of noble gas under an electric field (see Figure \ref{fig:TPC}). 

Electrons liberated in ionising interactions in the hydrogenous liquid volume, drift towards the liquid surface.  The electrons arriving at the surface must be collected with high gain and low noise. This can be achieved by a two step amplification:
\begin{enumerate}
\item Conversion of charge to light via electroluminescence with a noble gas, for example using xenon, since hundreds of photons can be produced by one drifting electron.
\item Conversion of light back to charge with the well known photomultiplier technology or with any replacement techniques currently being developed. Of course, the photon detector must be pixellised if tracking is desired.
\end{enumerate}
The idea is to measure precisely the charge liberated in interactions to obtain a good energy resolution. With pixellised photon detectors, a 3D image would be produced for each event, giving the ability to discriminate between interacting particles based on the spatial distribution of the energy depositions. Finely-grained images are needed to measure the direction of anti-neutrinos.

\subsection{Description of the Detection Technique}

Interacting charged particles ionise the neighbouring atoms along the track. Liberated electrons are drifted upwards through the liquid volume towards the noble gas layer. A strong electric field will drag the electrons in to the gas, where they will excite the gas molecules, producing VUV photons. The electroluminescence signal will be imaged by an array of photo-detectors.  The spatial and time structure of the recorded signals will give a 3D reconstruction of the event.  The total number of photons observed is directly proportional to the the number of drifted electrons and hence proportional to the energy of the event.

\subsubsection{Extending Known Technologies}
This detection method is based on known technologies.  
The use of organic liquids in ionisation chambers is not new \cite{cern, WALIC}. Many organic liquids ionise along the track of a charged particle, and can transport charge. An important advantage of these liquids is that they operate at room temperature and therefore do not require significant cooling unlike liquid argon or xenon. This is one reason for the interest in the late 70s to early 90s.  The liquids discovered are non-polar and most often have a spherical molecular structure. A critical step was the development of techniques to purify the liquids, removing electronegative impurities which can trap drifting electrons. Demonstrations of electron lifetimes of $\sim$ms were made, only possible with impurity levels of $\sim$ppb \cite{Engler1999}.  There are measurements of the $W_{i}$ and mobility for many liquids, which vary significantly from molecule to molecule. Table \ref{liquidproperties} gives  $W_{i}$ and mobilities for a selection of organic liquids.  For useful reviews see for example \cite{Engler1996}.

Extraction of the electrons from an organic liquid has been demonstrated in hexane and iso-octane (2,2,4-trimethylpentane) \cite{Bolozdynya1999,BBY1977}. Electrons arriving at the liquid surface can pass the potential barrier ($V_{0}$) if they have sufficient energy. This energy can be sufficient if the barrier is very low (or positive) for electrons at room temperature. For higher barriers, the electrons can acquire sufficient energy from the electric field (easily achieved in argon and xenon). Or the electrons can tunnel through the barrier. In this case, emission is expected to be slower, dependent on the electric field strength and liquid temperature. An important consideration in this latter case is the purity of the liquid; electrons can be trapped by electronegative impurities at the surface. The longer the time for extraction, the more chance of electron loss. Table \ref{liquidproperties} gives values for the potential barrier where known.  To our knowledge, no other attempts at electron extraction from organic liquids were made by any other groups.

Electroluminescence in noble gases is well studied and is at the heart of several projects aiming to detect Dark Matter \cite{LUX, Xenon, DARKSIDE} and Neutrinoless Double Beta Decay \cite{EXO}.  Traversing electrons excite gas molecules, which de-excite emitting VUV photons. The photon yield of electroluminescence is related to the gaseous electric field strength, gas density and is proportional to the drift distance. In two-phase liquid xenon detectors yields of $\sim$ 100 photons per drifting electron can be obtained such that a single extracted electron can be observed \cite{Edwards2007}.  

The energy resolution of the electroluminescence signal is limited by the number of free electrons produced in the initial interaction in the liquid.  Many materials, noble gases and liquids and semi-conductors have significant Fano factors which enhances their energy resolutions. On the assumption of no Fano factor, the energy resolution of the secondary signal would be limited by the Poisson statistics of the number of free electrons.
Measurements of the free electron yield in other organic liquids \cite{cern} indicate that $W_{i}$ values of $\sim 100 eV$ can be expected. This leads to approximately 10,000 electrons/MeV and a Poisson statistically limited energy resolution of 1\%.  Other factors then apply, such as the proportion of electrons extracted in to the gas phase, the fluctuation of the number of photons produced in the electroluminescence signal and the light collection.   

Operating at electric field strengths of $\sim$ 10 kV/cm and not substantially higher should allow an observable signal from electroluminescence without risking discharges or cracking of the organic vapour. The electroluminescence yield, from an admixture of noble gas and vapour from the organic liquid, must be measured and good operating parameters found. Liquids with low vapour pressures would be advantageous, see Table \ref{liquidproperties2} for some examples. Reducing the temperature slightly will have a large effect on the organic vapour pressure, most likely increasing the photon yield, and should be considered.

The idea of mixing organic liquids and noble gas vapours has been previously suggested \cite{Bolozdynya1999}. A small prototype comprising solid methane and gaseous neon was shown to work successfully.  Several liquid scintillators have also been shown to be capable of transporting charge \cite{Conkey2012}.

Tables \ref{liquidproperties} and  \ref{liquidproperties2} show some useful properties of a selection of previously studied organic liquids. This list is by no means exhaustive, for a more complete list of studied liquids see \cite{Holroyd}. Some of the organic liquids contain high Z atoms and so are interesting for gamma ray detection. In Table \ref{liquidproperties2} some interesting physical properties are shown which are worth considering for a large volume neutrino detector. The liquid density, hydrogen content, and hydrogen-to-carbon ratio are compared to the Double Chooz target scintillator cocktail and, the common scintillator component, pseudocumene. These liquids have comparable densities and the hydrogen content are similar. This is true even for the molecules containing high A atoms, for instance tetramethylgermane, which is surprising at first glance.  The reason is clear when considering the hydrogen-to-carbon ratio of these liquids, the proportion of carbon is much higher in liquid scintillators. In terms of the rate of neutrino interactions via Inverse Beta Decay on hydrogen, there is no gross disadvantage to having a high A atom in the molecule.  We also note that for a solar neutrino detector it would be advantageous to reduce the carbon content of the liquid, since $^{14}$C is the dominant background to $\nu_{e}$-electron scattering at low energies necessary to detect pp neutrinos. 

To explore the physics potential of such a TPC, we consider the organic liquid tetramethylgermane (TMGe) and xenon.

\begin{table}[h]
   \caption{\label{liquidproperties} Table of charge transport properties for a selection of organic liquids. Properties for liquid argon are shown as a comparison. $^{*}$ indicates liquids where electron extraction into the gas phase has been observed.}
   \begin{tabular}{|l|l|l|l|l|} 
     \hline
      {Liquid} & {W$_{i}$} & {Mobility $\mu_0$} & {V$_{0}$} & {Refs}\\
      {} & {(electrons/100eV)} & {($cm^2 V^{-1} s^{-1}$)} & {(eV)} &{}\\ \hline
      {n-hexane} & {0.13} &{0.09} & {+0.09 $^{*}$} & {\cite{Bolozdynya1999,Holroyd}} \\
      {2,2,4-trimethylpentane} & {0.33} & {6.6} & {-0.18 $^{*}$} & {\cite{Bolozdynya1999,Holroyd}} \\
      {2,2,4,4-tetramethylpentane} & {0.74} & {24} &{-0.3} & {\cite{Bolozdynya1999, Holroyd}} \\
      {tetramethylsilane} & {0.7}  & {100} & {-0.62}  &{\cite{Holroyd}}\\
      {tetramethylgermane} & {0.63} & {90} & {unknown} &{\cite{Holroyd}}\\
      {argon} & {4.23}& {475} & {-0.21 $^{*}$} & {\cite{Bolozdynya1999, Miyajima}} \\ \hline
   \end{tabular}
\end{table}

\begin{table}
   \caption{\label{liquidproperties2} Table of physical properties for a selection of organic liquids. Properties for the Double Chooz Gadolinium loaded scintillator cocktail are shown as a comparison.}
 \begin{tabular}{|l|l|l|l|l|} 
 \hline
  {Liquid} & {density} & {vapour pressure} & {H atoms} & {H/C ratio}\\
   {} & {(kg/m$^{3}$)} & {kPa} &{(10$^{25}$/kg)} & {} \\ \hline
   {n-hexane} & {0.655}  & {20.49}&{9.8}  & {2.33}  \\
   {2,2,4-trimethylpentane} & {0.69}&{5.5} & {9.5} & {2.25} \\
   {2,2,4,4-tetramethylpentane} & {0.72}& {2.67} & {9.4} & {3.33}  \\
   {tetramethylsilane} & {0.648}  & {74.65}& {8.2} & {3}  \\
   {tetramethylgermanium} & {0.978} & {46.46}& {5.45} & {3} \\
   {Double Chooz target scintillator} & {0.8035} & {N/A} &{8.12 \cite{DCScintillator}} & {$\sim$1.9}\\
   {pseudocumene} & {0.876} & {N/A} & {6.02} & {1.33} \\ \hline
 \end{tabular}
\end{table}

\section{Detecting Electron Anti-Neutrinos}
We consider the case of electron anti-neutrinos produced by a nuclear reactor.
The target considered is TMGe, with an interacting
anti-neutrino of energy greater than 1.8 MeV causing an inverse-beta decay of a free proton, to release a positron and a free neutron.
\begin{equation}
\bar\nu_e + p \rightarrow n + e^+ 
\end{equation}
As the interaction cross-section rises (with the square of the
energy) and the reactor neutrino spectrum falls in a similar fashion,
the convolution of these two, the observed spectrum is roughly
Gaussian in shape with a peak visible energy of $\sim$4~MeV, with no neutrinos emitted at energies higher than approximately 8 MeV.

In this reaction, the produced positron ionises the liquid and annihilates.  The annihilation gamma rays travel several centimetres, and Compton scatter. A clear advantage is to use a molecule with a high Z atom, to limit the number of Compton scatters the gamma rays undergo, then the identification of each annihilation gamma is simpler. This addition of a high Z component is a reason to favour materials like TMGe and is markedly different from all current liquid scintillator experiments.

The emitted neutron scatters and slows in the target material, most likely via elastic scattering on hydrogen. Once sufficiently slowed, the neutron can then be captured on hydrogen (releasing 2.2 MeV gamma rays), germanium (releasing $\sim$ 6 MeV in cascade gamma rays) or on a dopant isotope. The characteristic capture time is dependent on the material but can be expected to be of the order of $\sim$100s of $\mu$s for most hydrogenous materials. The addition of isotopes with higher neutron capture cross-sections than hydrogen speeds up the capture.

If the entire event is imaged in 3D. The neutrino energy can be estimated from the positron energy deposition alone: $E_\nu = E_{positron} + 1.8$ MeV. Positron tracks should be clearly visible for the majority of the neutrino spectrum (greater than $\sim$2 MeV).

Clusters of energy deposits from the emitted gamma rays (annihilation gamma rays and any gamma rays from neutron capture) should be observed. The total energy of which is known.

Identifying the presence of a neutron can also be made by doping the liquid with $^{10}$B or $^6$Li.  The advantage is that these isotopes emit alpha particles, giving a clear (and simple) marker to the presence of a neutron.  

The isotope $^{10}$B has a 20\% natural abundance and a thermal neutron capture cross-section of 3835 barns (remembering Hydrogen cross-section is 0.3 barns). The neutron capture cross-section is also favourable at epi-thermal neutron energies, permitting capture before thermalisation.

\begin{align}
n + ^{10}B \rightarrow &^{7}Li^{*} + \alpha   \mbox{ \hspace{70pt}($BR=94\%$, $Q=2.3MeV$)} \\
   &^{7}Li^{*} \rightarrow ^{7}Li + \gamma \mbox{ \hspace{40pt}($E_{\gamma}=0.48MeV$)} \nonumber \\
\nonumber \\
n + ^{10}B \rightarrow &^{7}Li + \alpha  \mbox{ \hspace{72pt}($BR=6\%$, $Q=2.8MeV$)}
\end{align}

SRIM calculations show that the alpha track length one can expect is only 24$\mu$m, ie the interaction is effectively point-like, and so quenching is expected to be strong.  There are very few measurements of $W_i$ for alpha particles in organic liquids, so this must be measured.  For  $^{10}$B there is often a de-excitation gamma ray.

\section{Electron Anti-Neutrino Directionality}
The angular correlation between the antineutrino direction and the initial direction of the neutron is described in \cite{vogel}. At threshold the neutron is emitted purely forwards. As the antineutrino energy increases, the average angle between the anti-neutrino direction and that of the neutron emission increases. For energies below 4.5 MeV, the maximum angle obtainable is still below 45$^{\circ}$, and for the highest reactor energies the maximum angle is $\sim$ 60$^{\circ}$.

To obtain the best estimation of the direction of the incoming anti-neutrinos, the interaction origin of the positron and the initial neutron direction must be well measured.  For the positron, one expects to measure the whole track, and this track will not be straight. However, identifying the end point (identified by the Bragg peak) and tracing back to the start point should be possible.

In most applications, the neutron capture point is used to determine the emitted neutron direction. However, the angular resolution obtainable with this technique is limited due to the multiple elastic scatterings of the neutron on hydrogen before capture.  The hydrogen density of the liquid effectively determines the average displacement of the neutron before capture and the ratio between the neutron capture species and Hydrogen the angular resolution. One idea is to dope with isotopes such as $^{10}$B, to favour early epi-thermal neutron capture.  However, large dopings are required such this may be impractical.  

Another method is to identify the first proton recoil produced by the emitted neutron. If correctly identified, this gives a perfect direction vector of the neutron.  This can be attempted by scanning the 3D image to search for a proton recoil close to the positron track. Proton recoil track lengths are small, quenching can be expected to be high (similar to alpha particles), and the energy deposits will be of $\sim$ few keV. 

This can only be attempted due to the fact that the proton recoils appear on the image of the event. No triggering is required on these low energy signals.

Another advantage of this technique is that the proton recoil is effectively instantaneous with the positron.  Such that no matter the drift speed of the electrons, the time between the electroluminescence signals of the positron and proton recoil gives the z-displacement of the neutron.

\begin{figure}
\begin{center}
  \includegraphics[width=10cm]{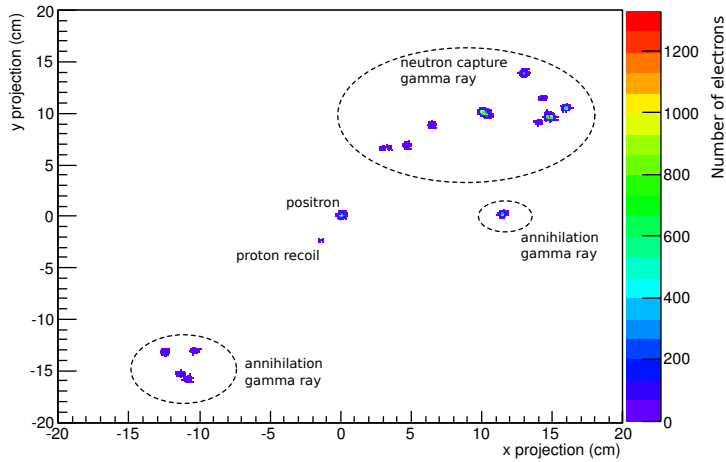}
\caption{\small Spatial distribution of electrons for a typical anti-neutrino interaction. In the centre, the positron and first proton recoil, at larger distances, the gamma ray interactions. The two 511 keV gamma rays are clearly identified as are the Compton scatters from a 5 MeV gamma ray emitted from the neutron capture on Ge.}
\label{fig:antineutrino_example}
\end{center}
\end{figure}

\begin{figure}
\begin{center}
  \includegraphics[width=10cm]{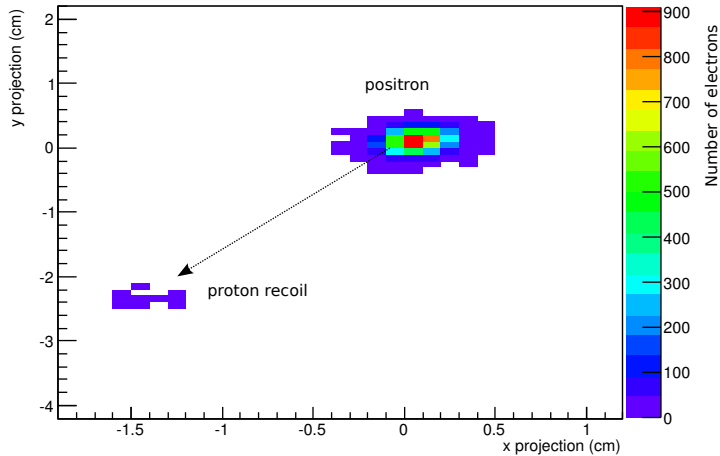}
\caption{\small Positron track and first proton recoil of emitted neutron. In this example, the first proton recoil occurs $\sim$ 2 cm from the anti-neutrino interaction point. There is a clear separation of the positron and proton recoil event, and the neutron direction is evident.}
\label{fig:antineutrino_example_zoom}
\end{center}
\end{figure}

A simple simulation was made to explore the topology of interactions produced by anti-neutrinos from the Chooz nuclear reactors.  GEANT 4 was used to track the particles through pure TMGe (no doping), giving the spatial co-ordinates of the energy deposits.  These energy deposits were converted into free electrons assuming a W$_{i}$ of 100 eV. For the nuclear recoil interactions, the number of electrons liberated was quenched assuming a quenching factor of 0.01. Diffusion on the drifting electrons was assumed to follow a Gaussian distribution with a sigma of 1 mm. Figures \ref{fig:antineutrino_example} and \ref{fig:antineutrino_example_zoom} show the projection onto the liquid surface of the liberated electrons, sampling at 1 mm, for a typical event.  If the electroluminescence and optical system is well controlled, these tracks are representative of the images that could be obtained from the TPC.

\section{Conclusion}
We describe the concept of a organic liquid TPC.  We show how such a detector could improve upon existing anti-neutrino detectors by meeting three challenges: 
\begin{itemize}
\item Good energy resolution (goal $<3\%$ at 3 MeV) - obtained by measuring charge.
\item Discrimination between positrons and betas - requires spatial resolution of $\sim$cm 
\item Measuring the direction of anti-neutrinos - requires spatial resolution of $\sim$mm 
\end{itemize}

Further simulation work is required in order to determine the parameters of the optical system required and to set bounds on the acceptable limits of parameters such as the drift speed, longitudinal and transversal diffusions, and gas depth.

A strategy is needed to analyse the resulting images, with estimations of the efficiency of correctly identifying the positron, proton recoil and events from the Compton scattered gamma rays.  It is also possible that there are other physics processes that can mimic the anti-neutrino signatures, this must be explored.
 
Recently a liquid scintillator TPC was proposed as an alternative to liquid noble gas detectors for neutrino physics \cite{Conkey2012}. Here the focus is on a hydrogenous-TPC for anti-neutrino detection, with origins at reactors or for geo-neutrinos.  However, similarly to the liquid scintillator TPC, there is potential for other applications, in particular the search for Neutrinoless Double Beta decay.  The first liquid considered, TMGe, contains the well-known double beta isotope $^{76}$Ge. Using the topology of the event, the signal, comprising 2 electrons with the same origin, would be completely separable from other background interactions. The energy resolution of the detector ($\sim 3\%$) would be comparable to other Double Deta Decay experiments. Other liquids containing Double Beta isotopes could also be used, for example tetramethyltin. 

An interesting feature of this proposed technology is the variety of potential liquids that could fill the same detector. This allows a wide range of potential physics goals. Unlike the liquified noble gas chambers, a cocktail of organic liquids could be used to obtain desired quantities such as hydrogen density, doping with particularly useful or interesting atoms, tuning the response to gamma rays etc.

A significant advantage of this TPC is that it operates at room temperature. Other general advantages which could be useful for other applications include:
\begin{itemize}
\item Good energy resolution 
\item Tracking
\item Good gamma-ray and neutron detection efficiency
\item Large active volume
\end{itemize}

A demonstration of electron extraction, electroluminescence, and subsequent measurements of the electroluminescent yield, with a mixture of organic vapour and noble gas is needed to show the feasibility of the TPC.  There is a potentially interesting technology to develop which could have impact in many other fields.

\subsection*{Acknowledgments}
We would like to thank Herve de Kerret for insightful suggestions in particular regarding the measurement of neutrino direction.  We are very grateful to Alexander Bolozdynya for his advice on organic liquid TPCs (emission detectors).
We thank John Losecco, Luis Fernandez Gonzalez and Lindley Winslow for their comments. We thank also Davide Franco for providing us with a GEANT 4 detector simulation.

\end{document}